\documentclass[10pt,conference]{IEEEtran}
\usepackage[utf8]{inputenc}
\usepackage{amsmath,amssymb}
\usepackage{booktabs}
\usepackage{graphicx}
\usepackage{multirow}
\usepackage{url}
\usepackage[hidelinks]{hyperref}
\usepackage{microtype}
\usepackage{balance}

\begin{document}

\title{The Defect Is in the Arithmetic Between the Files:\\
Static Detection of Exactly-Once Violations in Payment Pipelines}

\author{\IEEEauthorblockN{Abhishek Sharma}
\IEEEauthorblockA{Senior Member, IEEE\\
\texttt{abhicse24@gmail.com} \quad ORCID: 0009-0007-1103-2103}}

\maketitle

\begin{abstract}
Whether a streaming payment pipeline charges a customer twice after a restart
turns on a comparison nobody performs: how long the provider remembers an
idempotency key, against how stale a record can be when the pipeline processes
it for the second time. The first number is in the provider's documentation.
The second is not written anywhere. It is assembled from a checkpoint interval
in one file, a recovery assumption in a runbook, a restart delay in a third
and a source's lateness bound in a fourth. Each file is defensible alone; the
fault is in the relation between them. We give a checker that gathers those
facts and states the relations, and a synthesiser that turns 335 of the 515
findings into a runnable scenario: these records, kill the job here, bring it
back after this long, and this is the observable. 245 of them produce two
charges. The central
measurement is how far apart the evidence lies. Across a corpus of 630
pipelines we generated, 210 of them cannot be derived from fewer than
three artifacts, and no single artifact yields more than 35 of the 305
duration findings. Executing the witnesses corrected our own rules: two were
demoted after their scenarios produced a repeated call the provider's
idempotency absorbed rather than a second charge. All 16 selected live
witnesses reproduced their predicted observable against a real Kafka broker
with a real process kill, and a matched control in which the provider never
forgets the key did not, attributing the second charge to the key lifetime,
not the replay. Fault injection finds the same defects only if
the tester already assumed a long enough outage: one willing to simulate five
minutes finds 15\% of them, and 12 hours is needed for all. A sweep also
locates a configuration with no application-level fix, where above a 14-minute
recovery assumption the transaction timeout required to survive it exceeds the
broker's own cap.
\end{abstract}


\section{Introduction}

A payment pipeline reads orders, calls a provider to move money, and writes to
a ledger. It is configured for exactly-once: a transactional sink, a
deduplicating operator, an idempotency key on every call. Then the job dies at
lunchtime, somebody notices at two, and when it comes back it reprocesses
everything since its last checkpoint. Some of those customers are charged
twice.

Nothing in the pipeline is obviously wrong. The key is derived from the order
id, so the retry is recognisable. The provider does deduplicate. The sink is
transactional. What went wrong is that the provider stopped remembering the key
some hours before the replay arrived, and the replay was therefore a new
request as far as it was concerned.

That defect is a comparison between two numbers. One is in the provider's API
documentation. The other does not exist in any file: it is the longest a record
can go between its first processing and its second, and computing it needs a
checkpoint interval from the engine config, an assumption about how long a job
may stay dead from a runbook, a restart delay from a third place, and a
source's lateness bound from a fourth. No reviewer holds four artifacts at
once. No linter reads the provider's documentation. The pipeline passes every
review it will ever get.

This paper is about finding those before the outage. The contribution is in
three parts.

A \textbf{model that spans artifacts.} Facts are gathered from the engine
configuration, the client properties, the job SQL, the runbook and a
declaration of what the provider guarantees, and each fact keeps a record of
where it came from. From them we derive the \emph{replay horizon}, and state
the defect classes as relations against it. Thirteen checks, split between
structural defects a careful reviewer could find and duration defects that
require the arithmetic.

\textbf{Witnesses.} A static finding about a payment pipeline is easy to dismiss, so
335 of the 515 become a runnable scenario with a predicted observable --- the
rest describe a condition a single-record scenario cannot exhibit. Running them
against a real Kafka broker, with the pipeline process really killed and a stub
provider that really expires keys, either produces the predicted observable or
does not. For 245 of the 335 that observable is a second charge; for the
others it is a repeated call the provider's idempotency absorbs.
Every run is paired with a control in which the provider never forgets, so the
second charge is attributable to the key lifetime rather than to the replay.

\textbf{An evaluation that corrected the thing being evaluated.} Two of our rules
predicted a double charge and were wrong: their witnesses reproduce a repeated
call that the provider absorbs. Both were demoted to medium severity and their
witnesses now predict two invocations and one charge. The corpus's own clean
baseline turned out to be defective under a rule added later, which is reported
rather than quietly fixed.

One finding has no fix at the pipeline level. Kafka caps a producer's requested
transaction timeout at fifteen minutes by default, and surviving a recovery
assumption longer than about fourteen minutes requires more than that. The
choice is then between a shorter assumption, a cluster-wide broker change,
and giving up the transactional sink. None of the three is a setting in the
job.

\section{Background and related work}

\subsection{Exactly-once in stream processors}

Flink's checkpointing and Kafka's transactional producer give end-to-end
exactly-once for effects inside the system's own boundary: a record consumed,
transformed and written back to Kafka can be made to appear once. The
mechanisms are well documented --- asynchronous barrier snapshotting for
operator state~\cite{carbone2015}, and idempotent and transactional writes over
the log in Kafka~\cite{wang2021} --- and we assume rather than restate them.
Both are explicit about the boundary they hold within: Wang et
al.~\cite{wang2021} frame the guarantee over writes back into the log.

An external payment call is outside that boundary. The money moves before the
checkpoint completes, no two-phase commit can take it back, and the only
defence is the provider's idempotency. Everything in this paper is about that
gap, which is where payment pipelines actually live. Our contribution is not a
new exactly-once mechanism; it is a check on whether the durations a deployment
is configured with let the mechanism it already has do its job.

\subsection{Fault recovery and stream benchmarking}

Vogel et al.~\cite{vogel2024} benchmark fault recovery across stream processing
frameworks, measuring recovery time and throughput after failure. That is the
behaviour our duration checks depend on, measured empirically; what it does not
do, and does not set out to do, is decide whether a given configuration is
correct. Mohammad~\cite{mohammad2025} surveys Kafka design patterns and
benchmarking practice across 42 studies, finding recurring architectural
patterns; again a description of what people build rather than a check on it.

Fault injection searches the same space from the other end. Lineage-driven
fault injection~\cite{alvaro2015} reasons backwards from a correct outcome to
the failures that could have prevented it, and finds counterexamples in far
fewer executions than random injection. The faults it reasons about are message
and node failures whose effect is visible within the system under test.
Section~\ref{sec:results} measures what happens when the trigger is instead the
\emph{duration} of an outage: the tester has to sample a long enough downtime
before anything is observable at all, and our measurement is of how long that
has to be.

\subsection{Idempotency as a contract}

Idempotency keys are ubiquitous in payment APIs, and the mechanism is being
standardised: the IETF \texttt{Idempotency-Key} header
draft~\cite{idempotencykey} has the client supply a key, the server recognise a
repeat and return the original outcome. That specification is explicit that a
server retains a key for a finite period and that the client must not assume
retention beyond it, but it leaves the retention period to the server and says
nothing about how a client should establish that its own retries fit inside
whatever period it is given. That is the gap this paper addresses: the question
is not whether a key is supplied but whether the longest delay the deployment
can impose between two attempts fits inside the lifetime the provider offers.
We take the identity of an operation --- which two requests count as the same
money movement --- as given by the key the client supplies, and reason only
about durations.

\subsection{Static analysis and witnesses}

Reporting a defect together with evidence that it is real is an old idea in
verification. Beyer et al.~\cite{beyer2015} define an exchange format for
violation witnesses so that one verifier's counterexample can be replayed and
confirmed by another, precisely because a verdict without a checkable artifact
asks to be taken on trust. Our witnesses are of the same violation kind, with
one difference that matters here: theirs are validated by a second analyser
over the same program semantics, ours by execution against a running broker and
a provider that really expires keys. The defect's observable --- money moved
twice --- is decided by the provider's state rather than by agreement between
two tools, so execution is available to us as a check and we use it.

Static analysis that spans more than one artifact is likewise established
outside this setting. PCheck~\cite{xu2016} generates checks that catch
configuration errors at startup instead of at first use; cDep~\cite{chen2020}
recovers dependencies between configuration parameters within and across
components; Sotiropoulos et al.~\cite{sotiropoulos2020} detect missing ordering
and notification dependencies in Puppet by relating a program's dependency
graph to a recorded system-call trace. All three establish that correctness can
hinge on relations no single file states. What distinguishes the relations here
is their content: they are inequalities between \emph{durations}, one of whose
terms is not in the repository at all but in a third party's API
documentation, and another of which is an operational assumption written in
prose in a runbook.

What we did not find, and searched for, is a static checker for the
configuration relations below. Practitioner writing states several of them as
advice: keep the transaction timeout above the checkpoint interval plus
expected downtime; never expire an idempotency key before the client's longest
retry window. The advice is correct and nothing checks it.

\subsection{Sibling work}

ReconStream~\cite{reconstream} gives a temporal correctness contract for
streaming reconciliation, in which a verdict is admissible only once a source's
declared lateness bound has passed. Its notion of a declared bound is what the
replay horizon here borrows for its source term, and the reconciliation shape
in our corpus is its setting. The two share a shape of argument --- a
correctness property that relates durations rather than states --- applied at
different points of the pipeline. The defect families checked here are defined
in Section~\ref{sec:results}'s corpus and in Table~\ref{tab:rules}, and are not
inherited from elsewhere.

\section{The replay horizon}

Write $\mathrm{cp}$ for the checkpoint interval, $\mathrm{pause}$ for the
minimum pause between checkpoints, $\mathrm{down}$ for the longest a job is
assumed to stay dead, $\mathrm{restart}$ for the restart delay, and
$\Lambda_s$ for the declared lateness bound of source $s$. On recovery the job
resumes from its last checkpoint, so everything processed since is replayed;
while it was down the input accumulated, and draining that backlog takes time
proportional to the outage. Records still in flight when the job died arrive
within their source's bound. Putting those together,
\begin{equation}
\label{eq:horizon}
\begin{split}
H = {} & \mathrm{cp} + \mathrm{pause} + \mathrm{down} + \mathrm{restart} \\
       & {} + \delta(\mathrm{down} + \mathrm{restart}) + \max_s \Lambda_s ,
\end{split}
\end{equation}
where $\delta$ is the drain factor, one by default. $H$ is the age a record can
have reached when it is processed for the second time.

Two of the six terms come from the engine's configuration, one from a runbook,
one from a job definition, and $\delta$ is an operational estimate. That
spread is the reason $H$ is not computed anywhere, and computing it is most of
what the checker does.

\subsection{Two families of defect}

\textbf{Structural} defects are about shape: an effect that no commit protocol
covers, a key built from a fresh UUID, a consumer reading uncommitted output of
a transactional producer, an effect placed upstream of the deduplication meant
to guard it. A careful reviewer reading the job can find these.

\textbf{Duration} defects are relations against $H$. The provider forgets the
idempotency key before the replay arrives ($\mathrm{ttl} < H$); the dedup
state expires before the duplicate it exists to catch ($\mathrm{state\ ttl} <
H$); the broker aborts the transaction before the job returns; the client's own
retry window outlives the key. These are the ones that survive review, and they
are arithmetic.

Table~\ref{tab:rules} lists the thirteen checks. Each is a sufficient condition
for a violation, not a certificate of correctness. A pipeline with no findings has not been
proved exactly-once, and we are careful not to say otherwise.

\begin{table}[t]
\caption{The checks. ``Artifacts'' is how many separate files the finding's
evidence spans.}
\label{tab:rules}
\centering
\footnotesize\setlength{\tabcolsep}{5pt}
\begin{tabular}{llr}
\toprule
& & artifacts \\
\midrule
\multicolumn{3}{l}{\emph{structural}} \\
SC001 & effect outside the commit protocol & 1 \\
SC002 & non-deterministic idempotency key & 1 \\
SC003 & missing idempotency key & 1 \\
SC004 & uncommitted read of transactional output & 1 \\
SC005 & shared transactional id prefix & 2 \\
SC006 & effect upstream of deduplication & 1 \\
SC007 & non-idempotent sink under at-least-once & 1 \\
\midrule
\multicolumn{3}{l}{\emph{duration}} \\
SC100 & undeclared provider key lifetime & 3 \\
SC101 & key lifetime below the replay horizon & 3 \\
SC102 & dedup state TTL below the replay horizon & 3 \\
SC103 & transaction timeout below the downtime & 3 \\
SC104 & delivery timeout above transaction timeout & 1 \\
SC105 & client retry window above key lifetime & 1 \\
SC106 & required transaction timeout exceeds the broker cap & 2 \\
\bottomrule
\end{tabular}
\end{table}

\subsection{Reading the facts out of the files}

Four front-ends, one per place a fact lives. \texttt{flink-conf.yaml} gives the
checkpoint interval, the restart delay and the state TTL. The Kafka client
properties give the transaction and delivery timeouts and the consumer's
isolation level. The job SQL gives the topology: which tables are read, which are written,
whether a \texttt{ROW\_NUMBER} deduplication is present. The connector options
give the delivery guarantee and any watermark lateness. A
runbook gives the recovery assumption, if one is written down.

The fourth is different. How long the provider remembers an idempotency key is
a fact about the provider: it is in their API reference, it is load-bearing for
correctness, and no artifact in the repository records it. We ask for it in a
small manifest --- one declared duration per provider, alongside the key the
client already supplies~\cite{idempotencykey} --- and when it is absent the
checker emits a finding (SC100) rather than assuming a value.
Guessing would mean asserting something about a third party on no basis, and
the assumption would then be invisible in exactly the way the rest of the paper
is complaining about.

The SQL reader is deliberately shallow: it recognises the shapes these
pipelines are written in and reports what it cannot place, so a fact it fails
to find becomes a statement about incomplete evidence rather than a silent
default. Each extracted fact carries its origin, which is what lets a finding
cite the files it depends on, and what makes the
measurement in Section~\ref{sec:artifacts} possible.

\subsection{A worked example}

The artifacts in the repository's \texttt{data/example-pipeline} are ordinary.
The engine checkpoints every 60 seconds and restarts after 30. The producer
sets \texttt{transaction.timeout.ms=900000}, which is Kafka's default. The
runbook records a 45-minute worst observed recovery, with a note about a
weekend page that went unanswered for 40 minutes. The provider documents a
24-hour idempotency key. Each number is defensible and none would be questioned
in review.

Equation~\eqref{eq:horizon} gives $H = 1.54$ hours, comfortably inside the
24-hour key. The transaction timeout is not: surviving a checkpoint plus 45
minutes of downtime plus the restart needs 46.5 minutes, and 15 minutes is
configured. Raising it does not help, which is the next section.

\section{The defect with no fix}

Kafka caps a producer's requested transaction timeout at
\texttt{transaction.max.timeout.ms}. We read the value from a running broker
rather than from documentation:
apache/kafka 3.9.0 reports 900{,}000\,ms, and its synonym chain marks it
\texttt{DEFAULT\_CONFIG}. The refusal is measured too, at the boundary. A Java
producer calling \texttt{initTransactions()} against that broker is granted
60{,}000\,ms and granted 900{,}000\,ms, and refused at 900{,}001\,ms with
\emph{the transaction timeout is larger than the maximum value allowed by the
broker}. One millisecond either side of the cap is the difference between a
pipeline that starts and one that does not.

So for an exactly-once pipeline, the requirement $\mathrm{cp} + \mathrm{down} +
\mathrm{restart} \le \text{cap}$ bounds the recovery assumption rather than the
configuration. With 60-second checkpointing and a 10-second restart delay, any assumption
above 13.8 minutes is unsatisfiable; the bound moves between 9.8 and 14.7
minutes across the checkpoint intervals we swept. Every one of those figures
is conditional on four things we chose or measured: the broker's default cap
of 900{,}000\,ms, the drain factor of one held throughout, the replay-horizon
formula of~\eqref{eq:horizon}, and the recovery model in which a job is
restarted after a bounded outage. A cluster with a raised cap has no such
threshold, and a drain factor above one lowers it.

This is not a misconfiguration. It is a configuration that does not exist, and
the remedies (shorten the recovery assumption, raise the cap cluster-wide,
or drop the transactional sink) are all decisions above the pipeline. We
report it as a distinct rule (SC106) from the ordinary too-small timeout
(SC103), because telling a team to raise a number they cannot raise is worse
than telling them nothing.

\section{Witnesses}

A finding becomes a scenario: the input records, a fault schedule, and a
predicted observable. For a key-lifetime defect the scenario is one order,
a failure after the charge and before the checkpoint, and an outage long enough
for the key to expire; the prediction is two charges where there should be one.

\textbf{Witnesses run compressed.} A replay horizon of hours cannot be waited
out once per finding, let alone hundreds of times. Every duration check is an
inequality between durations and nothing else, so multiplying every duration in
a witness by one positive factor leaves each comparison, and therefore the
defect, unchanged. Witnesses are synthesised in the pipeline's units and
executed at a fraction of them. That is an argument about our own method, so we
measure it: Section~\ref{sec:results} reports the verdict of every witness
across four orders of magnitude of compression.

\paragraph{One witness, end to end} The boundary case whose key lifetime sits
one second under the replay horizon produces this. Compressed so the lifetime
is 30 seconds, the witness says: one order, fail once the charge has gone
through, stay down for 30.0 seconds. The pipeline's own log, from the run and
from the sufficient-lifetime control, differs in one line:

\begin{center}
\scriptsize
\begin{tabular}{@{}l@{}}
\texttt{start pid=$p_1$ restored=0 keys} \\
\texttt{effect ord-witness-1 key=ord-witness-1:stable charged=True} \\
\texttt{start pid=$p_2$ restored=0 keys} \\
\texttt{effect ord-witness-1 key=ord-witness-1:stable \textbf{charged=True}} \\
\midrule
\texttt{...control, lifetime long enough to cover the replay:} \\
\texttt{effect ord-witness-1 key=ord-witness-1:stable \textbf{charged=False}} \\
\end{tabular}
\end{center}

The key is identical in both. The record is identical. The kill is identical.
Six runs cleared the scaled 30-second lifetime and ten did not, and the
tightest of the six cleared it by about a second. That margin is wall clock and
it moves between runs that change no code, by enough that we quote it as a
range in Table~\ref{tab:live} and make the claim about the ordering rather than
the value. What reproduces is which runs clear the lifetime: the same six every
time, and the other ten never, with the gap falling inside the lifetime in one
arm and outside it in the other. Nothing about the pipeline distinguishes the two runs,
which is the argument for why the defect is in the configuration rather than in
the code.

\textbf{Some findings get no witness.} A shared transactional id prefix needs a
second job to manifest; an undeclared key lifetime is the absence of a fact
rather than a fact. Saying a rule has no runnable scenario is better than
shipping one that demonstrates nothing, and the coverage figure in
Section~\ref{sec:results} counts the gap honestly.

\section{Corpus and setup}

Five pipeline shapes, drawn from things payments teams build: taking a capture
to a processor, issuing a refund, reconciling a ledger against a bank feed,
paying out a batch, retrying a failed charge. Twenty-one mutations, each
introducing exactly one defect, crossed with the shapes, twice per combination,
over three seeds: 630 pipelines, of which 150 are negatives.

The mutations named after the rule meant to catch them are the weakest part of
the corpus and we treat them as such. The same author wrote the mutation and
the check, so a high score there measures implementation correctness rather
than detection power. The cases that carry evidential weight are the other two
kinds. Four \emph{boundary} mutations put a lifetime one second either side of
the replay horizon, so distinguishing them requires computing $H$ correctly
from all six terms; a checker that recognised mutation names rather than doing
arithmetic would fail them whichever way it erred. Two \emph{decoys} look
alarming and are correct: two-second checkpointing, which shortens the horizon
and makes every relation easier to satisfy, and a long client retry window that
still fits inside the key's lifetime.

\begin{table}[t]
\caption{The corpus. Sixteen mutations name the rule meant to catch them and
are the weakest evidence; the boundary and decoy cases are the informative
ones.}
\label{tab:corpus}
\centering
\footnotesize
\begin{tabular}{llr}
\toprule
kind & correct answer & cases \\
\midrule
16 defect mutations       & the named rule & 480 \\
clean pipelines           & no finding     &  30 \\
boundary, defective side  & SC101 or SC102 &  60 \\
boundary, correct side    & no finding     &  60 \\
2 decoys                  & no finding     &  60 \\
\midrule
\multicolumn{2}{l}{5 shapes $\times$ 21 mutations $\times$ 2 $\times$ 3 seeds} & 630 \\
\bottomrule
\end{tabular}
\end{table}

The five shapes are a capture taken to a processor, a refund, a reconciliation
of a ledger against a bank feed, a batch payout, and a dunning retry. The
reconciliation shape is the setting of~\cite{reconstream} and carries its
source lateness bound,
which is why its replay horizon exceeds what the engine configuration alone
would suggest. It is the one place in the corpus where a fact from a job
definition changes a duration relation.

Execution uses apache/kafka 3.9.0 in KRaft mode, a pipeline process killed with
SIGKILL, and an HTTP provider that expires keys on a timer. Nothing about the
recovery is modelled: what is replayed is whatever the broker redelivers from
the last committed offset.

\section{Results}
\label{sec:results}

\subsection{Detection}

Over 630 pipelines the checker recovers every injected defect and reports
nothing on any of the 150 negatives, at a few microseconds per pipeline. The
recall figure should be read narrowly for the reason given above. The boundary
cases are the informative part: all 30 instances either side of each threshold
are classified correctly, on a two-second difference in configuration.

\begin{table}[t]
\caption{Detection over 630 pipelines. The boundary and decoy rows are the
informative ones; the mutation rows are matched to their rules by construction.}
\label{tab:detection}
\centering
\footnotesize
\begin{tabular}{lrrr}
\toprule
case kind & cases & correct & findings raised \\
\midrule
injected defects (16 mutations) & 480 & 480 & as expected \\
clean pipelines                 &  30 &  30 & none \\
boundary, defective side        &  60 &  60 & the expected rule \\
boundary, correct side          &  60 &  60 & none \\
decoys (look wrong, are right)  &  60 &  60 & none \\
\midrule
total                           & 630 & 630 & 0 false positives \\
\bottomrule
\end{tabular}
\end{table}

\subsection{How many files a finding needs}
\label{sec:artifacts}

\begin{figure}[t]
\centering
\includegraphics[width=0.85\columnwidth]{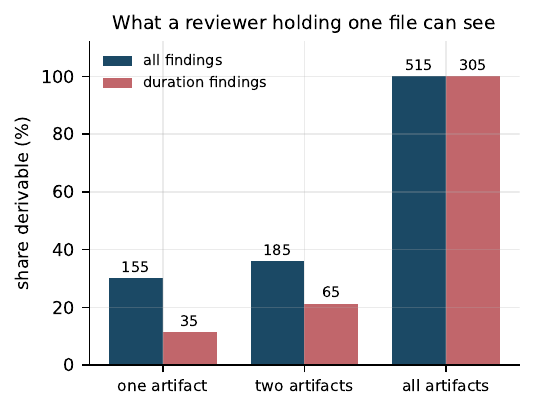}
\caption{Findings derivable from a restricted view of the artifacts.}
\label{fig:artifacts}
\end{figure}

Of 515 findings across the corpus, 245 are derivable from a single artifact, 60
need two, and 210 need three. Restricting to the 305 duration findings sharpens
it: 210 need three artifacts, 30 need two, and 65 need one. The two counts are
equal and duration findings are a subset of all findings, so the sets coincide:
\emph{every} finding that needs three artifacts is a duration finding. Needing
to read across the configuration is not a general property of static analysis
here --- it is what the replay horizon costs.

Figure~\ref{fig:artifacts} puts that in the reviewer's position. The most
informative single artifact yields 155 of 515 findings, and only 35 of the 305
duration findings, 11.5\% of them. The best pair reaches 185 findings and 65
duration findings. A reviewer reading the job, or the job and the producer
properties, sees the structural defects and almost none of the arithmetic. The
best-triple row of Table~\ref{tab:artifacts} carries the same count as the
single-artifact bucket of the distribution above, which is a coincidence of two
different measurements: one is what a chosen view can derive, the other is how
far apart a finding's evidence lies.

\begin{table}[t]
\caption{Findings derivable from a restricted view. ``Duration'' is the subset
that is arithmetic against the replay horizon.}
\label{tab:artifacts}
\centering
\footnotesize
\begin{tabular}{lrrrr}
\toprule
view & findings & share & duration & share \\
\midrule
best single artifact & 155 & 30.1\% & 35 & 11.5\% \\
best pair            & 185 & 35.9\% & 65 & 21.3\% \\
best triple          & 245 & 47.6\% & --- & --- \\
everything           & 515 & 100\%  & 305 & 100\% \\
\bottomrule
\end{tabular}
\end{table}

This is the paper's central measurement, and it is what the introduction's
claim rests on: these defects survive review because review is conducted one
file at a time.

\subsection{Do the witnesses reproduce?}

\begin{figure}[t]
\centering
\includegraphics[width=0.85\columnwidth]{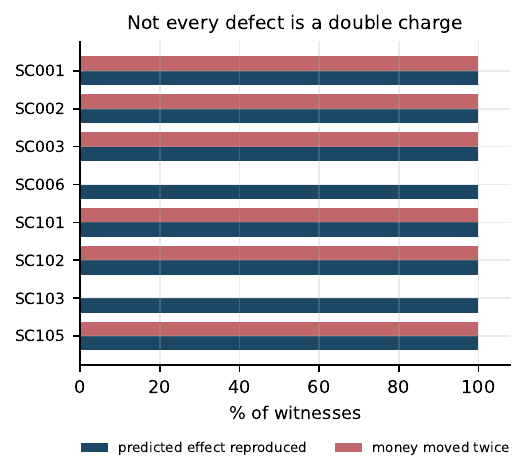}
\caption{Witness outcomes by rule. Reproducing the predicted effect and moving
money twice are different things.}
\label{fig:witness}
\end{figure}

Of 515 findings, 335 admit a witness (65\%; 70.5\% of the duration findings).
Every one of the 335 reproduces its predicted observable. Only 245 of them,
73\%, move money twice.

That gap is a correction to our own rules rather than a property of the
pipelines. Two rules originally predicted a double charge: an effect placed
upstream of its deduplication, and a transaction timeout too short to survive
the downtime. Both produce a second call to the provider, and with a sound key
still in its lifetime the provider absorbs it. The pipelines are wrong: they
have no defence of their own and are correct
only because the provider is. But they are not double-charging, and reporting
them as though they were would have been a false alarm on every one. Both were
demoted to medium severity and their witnesses now predict two invocations and
one charge.

\begin{table}[t]
\caption{Witness outcomes by rule. ``Reproduced'' is the predicted observable;
``charged twice'' is money actually moving a second time. Rules with no
witness describe conditions a single-record scenario cannot exhibit.}
\label{tab:witness}
\centering
\footnotesize
\begin{tabular}{lrrrr}
\toprule
rule & findings & witnessed & reproduced & charged twice \\
\midrule
SC001 & 30 & 30 & 100\% & 100\% \\
SC002 & 30 & 30 & 100\% & 100\% \\
SC003 & 30 & 30 & 100\% & 100\% \\
SC004 & 30 & --- & --- & --- \\
SC005 & 30 & --- & --- & --- \\
SC006 & 30 & 30 & 100\% & 0\% \\
SC007 & 30 & --- & --- & --- \\
SC100 & 30 & --- & --- & --- \\
SC101 & 60 & 60 & 100\% & 100\% \\
SC102 & 60 & 60 & 100\% & 100\% \\
SC103 & 60 & 60 & 100\% & 0\% \\
SC104 & 30 & --- & --- & --- \\
SC105 & 35 & 35 & 100\% & 100\% \\
SC106 & 30 & --- & --- & --- \\
\bottomrule
\end{tabular}
\end{table}

Six rules are sufficient on their own for a double charge: the three key
defects and the three lifetime defects. Table~\ref{tab:witness} and
Figure~\ref{fig:witness} separate them.

Compression does not change any verdict. All 114 witnesses give identical
results across factors spanning four orders of magnitude, which is what the
homogeneity argument predicts and is worth having measured rather than assumed.

\subsection{Against a real broker}
\label{sec:live}

Sixteen witnesses were replayed against apache/kafka 3.9.0, each in three
arms: as synthesised, with a provider that never forgets the key, and with a
provider whose key lifetime is finite but comfortably longer than the replay
takes. Forty-eight executions, no errors. All 16 selected live witnesses
reproduced their predicted observable.

What that arm does and does not exercise is worth stating plainly, because
``on a real broker'' can be read for more than it is. It uses a real Kafka
broker, a real process death by \texttt{SIGKILL} and a real consumer group
eviction. The provider is \emph{not} a commercial payment provider: it is a
local HTTP stub that issues charges against an idempotency key and expires that
key on a timer, which is the part of a provider's behaviour the rules reason
about. No commercial provider was contacted and no money moved; a ``charge'' is
a row in the stub's state. It does not use Flink:
there is no Flink checkpointing, no rescaling, and no transactional Flink sink
in this arm. The worker is a purpose-built consumer that commits offsets and
calls the provider, which is the smallest thing that can exhibit the
observable, and the sixteen witnesses are selected from the corpus rather than
being all of it. The claim is therefore about the predicted observables of
those sixteen, not about Flink's behaviour under recovery.

\begin{table}[t]
\caption{Live execution. ``inf'' is the control whose provider never forgets;
``suf'' is the control with a finite but sufficient lifetime. The gap is
between the two attempts, on the provider's clock, against a 30-second scaled
key lifetime. Each rule has two witnesses, which specify different downtimes;
that is what spreads the wider ranges, not measurement noise. Across the three
arms of one witness the gap moves by under a second. The charge columns do not
move at all.}
\label{tab:live}
\centering
\footnotesize
\begin{tabular}{llrrrr}
\toprule
rule & & gap (s) & def. & inf & suf \\
\midrule
SC101 & key lifetime below horizon & 30.6--47.9 & 2 & 1 & 1 \\
SC102 & state TTL below horizon & 30.6--32.4 & 2 & 1 & 1 \\
SC105 & retry window above lifetime & 30.6--48.0 & 2 & 1 & 1 \\
\midrule
SC001 & effect not idempotent & 5.8--6.2 & 2 & 2 & 2 \\
SC002 & non-deterministic key & 5.9--6.1 & 2 & 2 & 2 \\
SC003 & missing key & 5.8--6.1 & 2 & 2 & 2 \\
\midrule
SC006 & effect before dedup & 0.0 & 1 & 1 & 1 \\
SC103 & transaction timeout & 5.9--6.0 & 1 & 1 & 1 \\
\bottomrule
\end{tabular}
\end{table}

Table~\ref{tab:live} separates three behaviours, and the separation is the
result. The three lifetime defects charge twice as configured, once when the
provider never forgets, and once again when it forgets only after a lifetime
long enough to cover the replay. The observed gap exceeds the key lifetime in
exactly those six runs and in no other. That is the mechanism the rules claim,
observed rather than argued.

The three key defects charge twice in all three arms. This is correct and
worth stating: when the key is absent, or freshly generated on each attempt,
no lifetime helps, because the provider has nothing to recognise. Their second
charge is attributable to the key rather than to the lifetime, and the
experiment says so by the controls failing to remove it.

SC006 and SC103 charge once everywhere, agreeing with the interpreter.

\paragraph{An earlier version of this experiment was measuring the harness}
Witnesses are executed compressed, and the first live arm scaled key lifetimes
to 1.2 seconds. Killing the worker, waiting for the consumer group to evict it
and restarting costs about six seconds of real time, measured at 5.9 to 6.0
seconds across repetitions, so every key expired during the restart whatever
downtime the witness specified. Under that arrangement SC103
witnesses whose configured downtime was nineteen milliseconds reproduced a
double charge, which is not a result about SC103 but a result about our test
rig.

The tell was in the results table: a column showing the configured downtime
sitting three orders of magnitude below the lifetime it was supposed to
outlast. The fix has three parts, and all three are needed. Scale lifetimes to
thirty seconds, several times the overhead. Record the gap the provider
actually observed, so the comparison against the lifetime can be checked rather
than assumed. And add the sufficient-lifetime arm: if the rig were expiring
keys on its own, that arm would charge twice as well, and it charges once in
every one of the runs where it should.

We report this at length because the failure was silent in the direction that
flatters the paper. A harness that expires keys on its own confirms every
lifetime defect put to it.

\subsection{What testing would cost}

\begin{figure}[t]
\centering
\includegraphics[width=0.85\columnwidth]{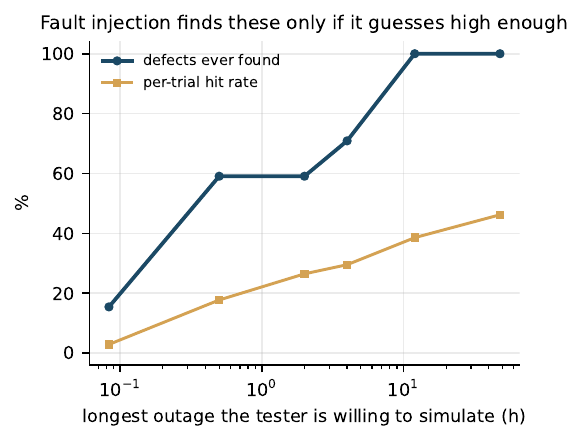}
\caption{A fault-injection tester finds these only if its assumed worst outage
is long enough. No number of trials substitutes for guessing high.}
\label{fig:testing}
\end{figure}

The alternative to reading the configuration is injecting faults and watching
for a double charge. It works, and Figure~\ref{fig:testing} shows what it
depends on. Our tester is given every advantage: it knows to kill the job, it
kills at the worst moment rather than a random one, and it inspects the
provider directly instead of waiting for a complaint. The only thing drawn at
random is how long the job stays down.

A tester whose longest simulated outage is five minutes finds 15\% of the
duration defects, and more trials do not help. The trigger is outside the distribution it
samples from. At thirty minutes it reaches 59\%, at four hours
71\%, and only at twelve hours does it find all of them. The per-trial hit rate
reaches 46.2\% at its highest, with a two-day ceiling.

\begin{table}[t]
\caption{Duration defects found by a fault-injection tester, as a function of
the longest outage it is willing to simulate.}
\label{tab:testing}
\centering
\footnotesize
\begin{tabular}{rrr}
\toprule
longest simulated outage & defects ever found & per-trial hit rate \\
\midrule
5 min  &  15\% &  2.9\% \\
30 min &  59\% & 17.8\% \\
2 h    &  59\% & 26.5\% \\
4 h    &  71\% & 29.5\% \\
12 h   & 100\% & 38.6\% \\
48 h   & 100\% & 46.2\% \\
\bottomrule
\end{tabular}
\end{table}

This is the honest form of the argument for static analysis here. It is not
that testing cannot find these defects. It is that testing finds them only if someone already suspected that the outage could be long enough, which
is the fact the checker computes, in microseconds, from files the team already
has.

\subsection{How much of the configuration space is affected}

\begin{figure}[t]
\centering
\includegraphics[width=0.85\columnwidth]{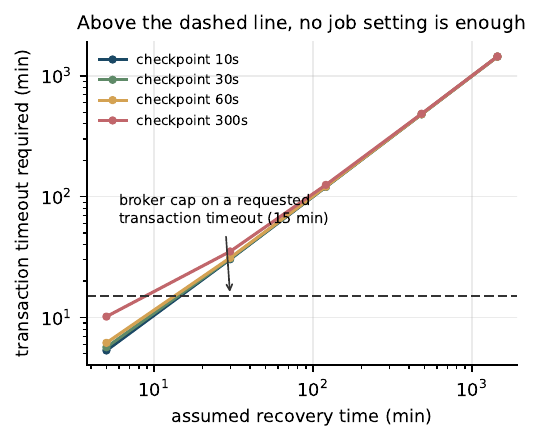}
\caption{Required transaction timeout against the assumed recovery time. The
dashed line is the broker's cap on what a producer may request.}
\label{fig:cap}
\end{figure}

Sweeping 960 combinations of checkpoint interval, recovery assumption, key lifetime, state TTL and transaction timeout, each drawn from values
practitioners use, 86.7\% produce at least one finding. That number is a
property of our menus and not of the field, and we do not present it as
prevalence. What transfers is the threshold in Figure~\ref{fig:cap}: the
required transaction timeout grows with the recovery assumption and crosses the
broker's cap at a recovery assumption of about fourteen minutes, after which no
configuration of the job is sufficient.

\section{Discussion}

\paragraph{The unit of review is wrong} Every artifact in the worked example
passes review. The pipeline does not. Code review, linting and configuration
validation are all organised around one file at a time, and the defects that
matter most here are precisely those that no file contains. A checker that
spans artifacts is a small thing technically and a different thing
organisationally.

\paragraph{One fact has no home} The provider's key lifetime is load-bearing
for correctness and lives in no repository. We ask for it to be declared
alongside the key the client already supplies~\cite{idempotencykey}, and treat
its absence as a finding rather than defaulting it. A tool that guessed would be asserting
something about a third party it has no basis for.

\paragraph{The horizon is an operational number, not an engineering one}
Four of the six terms in~\eqref{eq:horizon} come from configuration. The one
that dominates is not: how long a job may stay dead is a property of who is on
call, how alerts are routed, and whether the page goes out at three in the
afternoon or three in the morning. Shortening it is the cheapest fix for every
duration defect here, and it is the one no code change achieves. The tool's
most useful output may be that a number in a runbook is load-bearing for
correctness, which is not where teams look for correctness arguments.

\paragraph{Execution as a check on the checker} The most useful thing the
witnesses did was disagree with us. Two rules were wrong about their own
severity, and no amount of re-reading them would have shown it; running them
did, in seconds. We would recommend the arrangement to anyone writing a
checker whose findings have an observable consequence.

\section{Limitations}

The corpus is generated by us, and the mutations named after rules are matched
to those rules by construction. The boundary cases, the decoys and the live
execution are the parts that carry evidence; the headline recall number is not.

The execution arm is not Flink. Expressing an external HTTP effect in Flink SQL
requires a custom connector, and writing one would have put more of our own
code between the witness and the result. The replay semantics exercised are
Kafka's real offset commit and redelivery, and the process really dies, but
operator-level checkpointing and rescaling are not exercised. Pipelines are
single-partition throughout.

Both halves of the broker cap are now measured on one broker: its configured
value, and its refusal at one millisecond past that value. Neither is a
survey. A cluster whose operator has raised
\texttt{transaction.max.timeout.ms} has a different cap and a different
threshold, and we tested the default on apache/kafka 3.9.0 rather than a
distribution of deployed settings.

Each check is sufficient for a violation, not necessary. A pipeline with no
findings has not been proved correct, and the checker says so.

The provider in every arm is a local stub that implements an idempotency key
with an expiry. That is deliberate --- it is the one provider behaviour the
rules depend on, and it lets the key be expired on demand --- but it means the
live results are evidence about the defect families in Table~\ref{tab:rules}
under a provider that behaves as documented, not about any particular
commercial provider's implementation. Reproducing a witness shows that the
predicted observable follows from the configuration; it does not establish that
the thirteen checks are complete for exactly-once violations in general, and we
make no such claim.

The live arm covers sixteen witnesses, two per rule that admits one, rather
than the whole corpus: three arms at roughly a minute each puts an exhaustive
run beyond a reasonable budget. The interpreter covers every witness and agrees
with the broker on all sixteen, which is some comfort and not a substitute.

The drain factor, how long a backlog takes to clear relative to the outage
that produced it, is held at one throughout and never swept. A busier
pipeline drains more slowly and has a longer horizon than we compute, so our
estimate is optimistic in a direction that matters.

The configuration sweep uses menus we assembled from values practitioners
report. It supports a statement about where the threshold lies, not about how
common the defect is in deployed systems, and we have no field data.

\balance
\section{Conclusion}

The question of whether a restart charges a customer twice has an answer, and
the answer is arithmetic between numbers that live in four different files and
one external document. Gathering them takes microseconds and finding the
defects takes a comparison; what makes them survive is that nothing gathers
them. Two things surprised us in the building. Executing the witnesses demoted
two of our own rules, because a duplicate call the provider absorbs is not a
double charge. And one configuration turns out to have no fix at the pipeline level at all:
past a fourteen-minute recovery assumption, under the broker's default cap and
a modelled drain rate of one, the transaction timeout that would be needed is
larger than the broker will grant. The sweep locates that threshold; it does
not say how many deployed pipelines sit past it, and we make no estimate of
prevalence.

\section*{Artifact availability}

Code, corpus generator, Docker configuration and every results file are at
\url{https://github.com/abhisheksharma2411/streamcheck}, MIT licensed, at
release \texttt{v1.0.2}, which is the revision this paper was built from.
Releases are archived at \mbox{Zenodo} under the concept DOI
\mbox{\texttt{10.5281/zenodo.24412123}}, which resolves to the most recent
archived version.
\texttt{make reproduce} runs the pipeline
from a clean checkout: it starts a Kafka broker, runs every experiment,
regenerates each figure from the raw results and rebuilds this paper. No number
in the text is entered by hand; \texttt{scripts/verify\_claims.py} checks each
one against the results files.


\begin{thebibliography}{99}

\bibitem{vogel2024}
A.~Vogel, S.~Henning, E.~Perez-Wohlfeil, O.~Ertl, and R.~Rabiser,
``A comprehensive benchmarking analysis of fault recovery in stream processing
frameworks,'' in \emph{Proc. 18th ACM Int. Conf. Distributed and Event-Based
Systems (DEBS)}, 2024. doi:10.1145/3629104.3666040.

\bibitem{mohammad2025}
M.~Mohammad, ``Analysis of design patterns and benchmark practices in Apache
Kafka event-streaming systems,'' arXiv:2512.16146, Dec.~2025.

\bibitem{reconstream}
A.~Sharma, ``When Is a Reconciliation Break Real? A Temporal Correctness
Contract for Streaming Financial Reconciliation,'' arXiv:2610.14470, 2026.

\bibitem{carbone2015}
P.~Carbone, G.~F\'ora, S.~Ewen, S.~Haridi, and K.~Tzoumas, ``Lightweight
asynchronous snapshots for distributed dataflows,'' arXiv:1506.08603, 2015.

\bibitem{wang2021}
G.~Wang \emph{et al.}, ``Consistency and completeness: Rethinking distributed
stream processing in Apache Kafka,'' in \emph{Proc. ACM SIGMOD}, 2021,
pp.~2602--2613. doi:10.1145/3448016.3457556.

\bibitem{idempotencykey}
J.~Jena, S.~Dalal, and D.~Tishkin, ``The Idempotency-Key HTTP header field,''
IETF Internet-Draft \texttt{draft-ietf-httpapi-idempotency-key-header}, work
in progress, 2026.

\bibitem{xu2016}
T.~Xu, X.~Jin, P.~Huang, Y.~Zhou, S.~Lu, L.~Jin, and S.~Pasupathy, ``Early
detection of configuration errors to reduce failure damage,'' in \emph{Proc.
USENIX OSDI}, 2016, pp.~619--634.

\bibitem{chen2020}
Q.~Chen, T.~Wang, O.~Legunsen, S.~Li, and T.~Xu, ``Understanding and
discovering software configuration dependencies in cloud and datacenter
systems,'' in \emph{Proc. ACM ESEC/FSE}, 2020.
doi:10.1145/3368089.3409727.

\bibitem{sotiropoulos2020}
T.~Sotiropoulos, D.~Mitropoulos, and D.~Spinellis, ``Practical fault detection
in Puppet programs,'' in \emph{Proc. ACM/IEEE ICSE}, 2020, pp.~26--37.
doi:10.1145/3377811.3380384.

\bibitem{beyer2015}
D.~Beyer, M.~Dangl, D.~Dietsch, M.~Heizmann, and A.~Stahlbauer, ``Witness
validation and stepwise testification across software verifiers,'' in
\emph{Proc. ACM ESEC/FSE}, 2015, pp.~721--733. doi:10.1145/2786805.2786867.

\bibitem{alvaro2015}
P.~Alvaro, J.~Rosen, and J.~M.~Hellerstein, ``Lineage-driven fault
injection,'' in \emph{Proc. ACM SIGMOD}, 2015, pp.~331--346.
doi:10.1145/2723372.2723711.

\end{thebibliography}
\end{document}